\documentclass[rmp,twocolumn,preprintnumbers,superscript,amsmath,amssymb]{revtex4}
\usepackage{graphicx}
\usepackage{dcolumn}
\usepackage{bm}

\begin{document}


\title{Direct Observation of Local Magnetic Field Generated by Micro-Magnet }

\author{Shinji Watanabe}
\affiliation{Center for the Advancement of Higher Education,
Tohoku University, Sendai 980-8578, Japan}
\email{wshinji@mail.tains.tohoku.ac.jp}
\author{Susumu Sasaki, Shinya Sato}%
\affiliation{
Graduate School of Science and Technology, Niigata University, Niigata 950-2181, Japan}
\author{Naoki Isogai and Yoshinori Matsumoto}
\affiliation{Department of Applied Physics and Physico-Informatics, Keio University, Kanagawa 223-8522, Japan}

\begin{abstract}
Using standard nuclear magnetic resonance (NMR) technique and a well-fabricated sample, we have succeeded in directly observing local magnetic field generated by a micro-magnet Ni$_{45}$Fe$_{55}$ (the thickness of 400-nm) which was sputtered on an Al layer of 20-nm thickness. Improved sensitivity of our NMR technique enabled us to clearly observe Al-NMR signals, which are confirmed to come from Al nuclei in the 20-nm layers. From the analysis of the Al-NMR spectra, the local magnetic field was found to be +0.17$\pm$0.02 (-0.20$\pm$0.01) Tesla, the sign of which is consistent with the geometry that the external magnetic field was applied perpendicular (parallel) to the Al layer. The present study gives a potential key element toward realizing higher resolution in magnetic resonance imaging (MRI).
\end{abstract}

\maketitle

Based on the rapid progress of pulse techniques in NMR, magnetic resonance imaging (MRI) was invented and has been a well-known and well-established powerful method to obtain nondestructively three-dimensional (3D) images of human cells and organs\cite{1,2}. 
To obtain 3D image data, MRI requires a gradient in the static magnetic field. 
Since NMR frequency is exactly proportional to the magnetic field applied, the gradient results in   spread of frequencies.
Thus, the spatial resolution of MRI depends on not the magnitude but the gradient of the magnetic field. 
So far, the resolution of MRI systems for medical use is of the order of millimeters or micrometers\cite{3,4,5}. 
To obtain higher resolution, the sample should be smaller, and hence, the smaller amount of nuclei. 
This results in reducing signal to noise ratio (SNR), because NMR detects the nuclear-spin signals through coils ("inductive detection") placed around the material of concern. 
Contrary to these technical difficulties in MRI, it has become possible to detect a single \textit{electron} spin in magnetic resonance force microscopy (MRFM)\cite{6}, because a cantilever, instead of a coil, probes the spin signal\cite{7}. 
As a result, they have succeeded in achieving a detection sensitivity of roughly 1,200 \textit{nuclear} spins at 600 mK\cite{8}. 
Although the sensitivity of conventional MRI is much poorer than that of the MRFM, a wide variety of many pulses can easily be irradiated through the coils and a number of pulse sequences has been well-established. 
Thus, a breakthrough can be expected if the spatial resolution in inductive NMR is developed based on high-sensitivity technique. 
As can easily be seen from the principle of resonance, the greater the gradient in the magnetic field, the higher the resolution. 
One of the most practical ways to produce a greater gradient than that in standard MRI would be to place a micro-scale ferromagnet close to a thin layer of nuclei 
of concern.

In this Letter, we show that, employing inductive NMR method, we have succeeded in the detection of signals coming from the nuclei that feel the local magnetic field generated by a micro-scale ferromagnet. A well-fabricated sample and improved sensitivity of our NMR system enabled us to probe Al-NMR signals coming from a 20-nm Al layer above which a Ni$_{45}$Fe$_{55}$ alloy of 400-nm layer was sputtered. From the difference in the Al-NMR spectra under different orientations of the magnetic field, we confirmed that the Al signals surely stem from the Al-layer of 20-nm thickness that feels the local magnetic field. We also discuss the gradient of the local magnetic field from the simulation of the Al-NMR spectra.

Figure 1(a) shows a schematic view of our sample used in this study. 
The Ni$_{45}$Fe$_{55}$ of 400-nm thickness produces local magnetic field, $B_{local}$, in the Al-layer of 20-nm thickness just below. 
From a dc SQUID measurement, the magnetization of the sample was found to saturate when an external magnetic field $B_{ext}$ over 3 Tesla was applied. 
As can be seen from Fig. 2 (a) and (b), the Ni$_{45}$Fe$_{55}$ produces positive (negative) $B_{local}$ at the Al layer when the $B_{ext}$ is parallel to the $z$ ($y$) axis. 
As a reference for the Al-NMR spectra, we placed an Al metal [Al(ref)] in the NMR coil along with the sample. 
The coil was made of pure-Ag metal, which helped us not to observe the signal from the coil. 
We irradiated the sample with RF pulses of a given frequency, $f_{op}$, that the signal generator produces. 
As a response from the nuclei, we recorded the spin-echo intensity with sweeping the $B_{ext}$. 
All the Al-NMR spectra were obtained for the $B_{ext}$ over 3 Tesla to maximize the $B_{local}$. 
To detect the nuclear-spin signals in the 20-nm layer, we did the followings. 
First, we wound the coil directly on an insulating thin tape that wraps both the sample [10-mm width, 20-mm distance, 0.5-mm thickness (see Fig. 1(a)) $\times$ 3 pieces] and the Al(ref) [4-mm width, 12-mm distance, 10-$\mu$m thickness $\times$ 1 piece]. 
Second, using a network analyzer, we realized an ideal impedance matching of the resonance circuit; the Smith Chart showed that the imaginary part was $0 \pm 1$ ohm and the real part $50\pm 1$ ohm, and the amplitude of the reflection indicated $-50\pm10$ dB at around the $f_{op}$. Third, using a standard $^{4}$He-flow cryostat (Cryoindustry Co. Ltd), we cooled the sample down to 1.7 K. 
In reducing the $^{4}$He-gas pressure, we confirmed that the RF pulses did not result in arching. 
Fourth, in observing the spin-echo signal, we employed the quadrature detection to increase SNR, and phase-cycling techniques to cancel ring down noises caused by the coil \cite{9}.
\begin{figure}[t]
\includegraphics{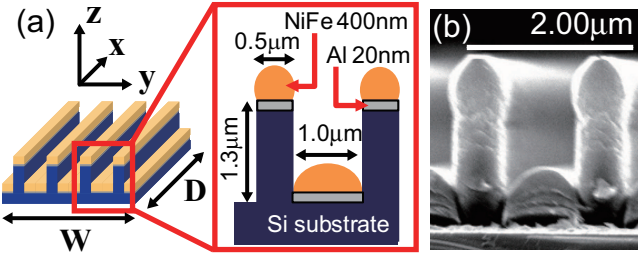}
\caption{\label{Fig.1} (Color online) (a): Schematic view of the sample. The width $W$ is 10 mm and the distance $D$ is 20 mm. The cross-section indicates the detailed structure. On the Al layer of 20-nm thickness, Ni$_{45}$Fe$_{55}$ alloy is sputtered with the thickness of 400 nm. (b): SEM image of the sample.
}
\end{figure}

We first obtained the NMR spectra with $B_{ext} \Vert z$ [Fig. 2(a)]. Figure 3(a) and (b) show the spectra obtained by sweeping the $B_{ext}$ at a given $f_{op}$. For simplicity and convenience, the $B_{ext}$ is shifted as $\frac{f_{op}}{^{27}\gamma}- B_{ext}$, where $^{27}\gamma$ (=11.094 MHz/T) is the gyromagnetic ratio of $^{27}$Al. A sharp peak observed nearly at the origin is assigned to Al(ref), which was confirmed by an experiment in which the Al(ref) was extracted from the coil. Here, we label the other signals as I, $\textrm{I\hspace{-.1em}I}$, $\textrm{$\textrm{I\hspace{-.1em}I}$\hspace{-.1em}I}$ and $\textrm{I\hspace{-.1em}V}$. In the followings, we clarify that Signal $\textrm{I\hspace{-.1em}V}$ comes from the 20-nm Al layer that feels the $B_{local}$, and that the others from materials outside the NMR coil.

First we show that Signal I and Signal $\textrm{I\hspace{-.1em}I}$ is assigned to $^{65}$Cu and $^{63}$Cu, respectively. Clearly seen for Signal I, the smaller the $f_{op}$, the smaller the $B_{ext}^{peak}$, where  $B_{ext}^{peak}$ is the peak position of the signal. It was found that $f_{op}=^{65}\gamma B_{ext}^{peak}$, where $^{65}\gamma$ (=12.089 MHz/T) is the gyromagnetic ratio of $^{65}$Cu. Thus, we can safely assign Signal I to $^{65}$Cu. In the same way, Signal $\textrm{I\hspace{-.1em}I}$ is assigned to $^{63}$Cu. This is validated by the fact that the intensity ratio of Signal I over Signal $\textrm{I\hspace{-.1em}I}$ is nearly equal to the natural abundance ratio of Cu isotopes, i.e., $^{65}$Cu : $^{63}$Cu = 30.9 \% : 69.1 \%. Here, it is quite natural to raise a question why Cu signals can be observed when the sample does not contain Cu nuclei and the coil is made of purely Ag. To clarify this, we performed measurements under the same conditions except that the sample was extracted from the coil. In the experiments, signals were observed at the peak positions of Signals I and $\textrm{I\hspace{-.1em}I}$. This indicates that these signals come from nuclei outside the coil. We speculate that they originate from a Cu tube which shields electromagnetic noises from outside, or a capacitor made of Cu metal. In general, this is not unusual but sometimes happens in high-sensitivity NMR measurements. In other words, the fact that we detected these signals proves the high sensitivity of the present measurements. In addition to Signals I and $\textrm{I\hspace{-.1em}I}$, Signal $\textrm{$\textrm{I\hspace{-.1em}I}$\hspace{-.1em}I}$ was observed in the experiment without the sample. Since the $B_{ext}^{peak}$ is close to the Al(ref), we speculate that Signal $\textrm{$\textrm{I\hspace{-.1em}I}$\hspace{-.1em}I}$ is ascribed to Al-NMR signal from Al$_2$O$_3$, the constituent of macor which our NMR probe contains.   

\begin{figure}
\includegraphics{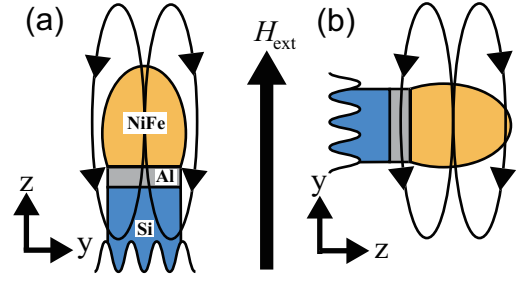}
\caption{\label{Fig.2} (Color online) (a): External field $B_{ext}$ applied in the $+z$ direction aligns the direction of magnetic domains in the micro-magnet, which produce additional \textit{positive} local field in the Al layer. (b): When $B_{ext}$ is in the $+y$ direction, \textit{negative} local field is produced in the Al layer.
}
\end{figure}
\begin{figure}[t]
\includegraphics{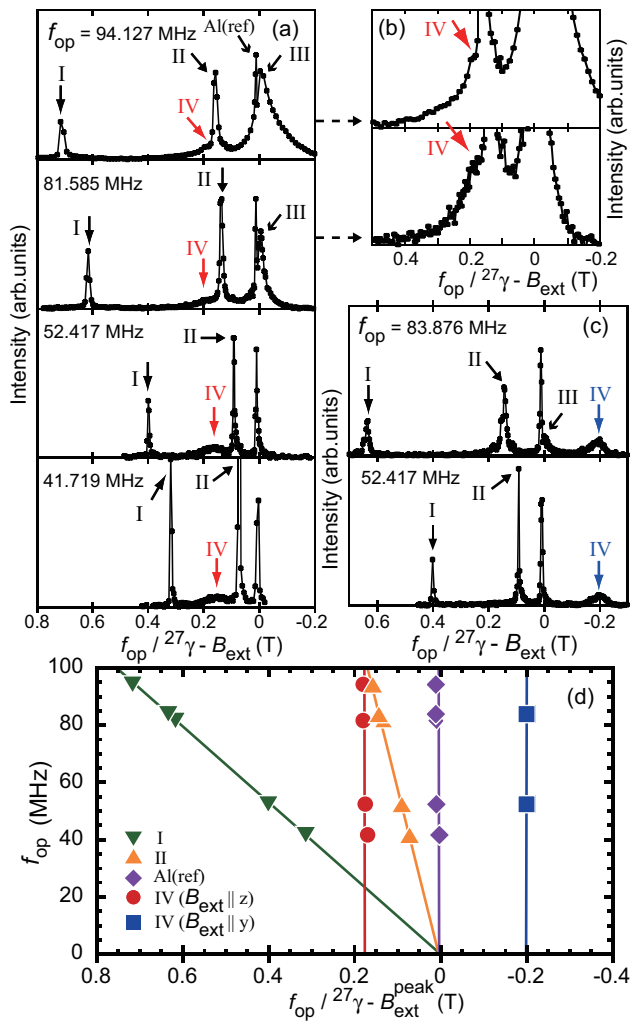}
\caption{\label{Fig.3} (Color online) Field-swept NMR spectra taken at different operational frequencies $f_{op}$ for $B_{ext} \Vert z$ (a) and for $B_{ext} \Vert y$ (c). To discuss Al peak positions, we express the external field as $\frac{f_{op}}{^{27}\gamma} - B_{ext}$. The Al(ref) peak is observed nearly at zero. Other signals are labeled as I, $\textrm{I\hspace{-.1em}I}$, $\textrm{$\textrm{I\hspace{-.1em}I}$\hspace{-.1em}I}$ and $\textrm{I\hspace{-.1em}V}$. (b): Expanded views around Signal $\textrm{I\hspace{-.1em}V}$ in the upper two of (a). (d): Peak positions, $B_{ext}^{peak}$, of Al(ref), Signals I, $\textrm{I\hspace{-.1em}I}$ and $\textrm{I\hspace{-.1em}V}$ plotted in the form of $f_{op}$ versus $\frac{f_{op}}{^{27}\gamma} - B_{ext}^{peak}$. All the errors are well within the marks. From the slopes of the fitted lines, we can assign Signals I, $\textrm{I\hspace{-.1em}I}$ and $\textrm{I\hspace{-.1em}V}$ as $^{65}$Cu, $^{63}$Cu and $^{27}$Al nuclei, respectively. Only Signal $\textrm{I\hspace{-.1em}V}$ does not cross the origin but is shifted by $+0.17\pm0.02$ Tesla for (a) and $-0.20\pm0.01$ Tesla for (c). Note that, for Al nuclei that feel the $B_{local}$, the relation that $f_{op}=^{27}\gamma(B_{ext}^{peak}+B_{local})$ should hold, which is equivalent to $B_{local}=\frac{f_{op}}{^{27}\gamma} - B_{ext}^{peak}$.
}
\end{figure}

Now we clarify that Signal $\textrm{I\hspace{-.1em}V}$ comes from the Al nuclei in the 20-nm layer. For the spectra at smaller $f_{op}$ (=52.417 MHz, 41.719 MHz), the peak position of Signal $\textrm{I\hspace{-.1em}V}$ is unambiguously defined, since the signals are clearly separated from Signal $\textrm{I\hspace{-.1em}I}$. On the other hand, this is not the case for the spectra at larger $f_{op}$ (=94.127 MHz, 81.585 MHz), because Signal $\textrm{$\textrm{I\hspace{-.1em}V}$}$ is superimposed on Signal $\textrm{I\hspace{-.1em}I}$. Thus, we defined the peak position of Signal $\textrm{I\hspace{-.1em}V}$ by the kink which can be more clearly seen in the expanded views [Fig. 3(b)]. In Fig. 3(d), we plot the peak positions of Al(ref), Signals I, $\textrm{I\hspace{-.1em}I}$ and $\textrm{I\hspace{-.1em}V}$ in the form of $f_{op}$ versus $\frac{f_{op}}{^{27}\gamma} - B_{ext}$. Except for Signal $\textrm{I\hspace{-.1em}V}$, due to the resonance condition that $f_{op}=\gamma B_{ext}^{peak}$, the fitted lines cross the origin and the slopes give the value of $\gamma[\frac{\gamma}{^{27}\gamma}-1]^{-1}$. In contrast, for Signal $\textrm{I\hspace{-.1em}V}$, the fitted line does not cross the origin but is shifted by $+0.17\pm0.02$ Tesla. It is to be noted that for the Al signals that feel the $B_{local}$, the values of $\frac{f_{op}}{^{27}\gamma} - B_{ext}^{peak}$(=$B_{local}$) should be non-zero and constant regardless of the $f_{op}$ values, as easily seen from $f_{op}=^{27}\gamma(B_{ext}^{peak}+B_{local})$. Thus, it is very likely to say that Signal $\textrm{I\hspace{-.1em}V}$ should come from the 20-nm Al layer that feels $B_{local}=+0.17\pm0.02$ Tesla.

To confirm this, we utilized the fact that the Al-layer should feel \textit{negative} $B_{local}$ if the $B_{ext}$ is applied in the $y$ direction [Fig. 2(b)]. 
In this case, the peak position of Signal $\textrm{I\hspace{-.1em}V}$ should be opposite in reference to the peak of the Al(ref). 
This is exactly what Fig. 3(c) shows, as expected. 
We would like to stress that, as is illustrated in Fig. 3(d), the values of $\frac{f_{op}}{^{27}\gamma} - B_{ext}^{peak}$(=$B_{local}$) are shifted by $-0.20\pm0.01$ Tesla irrespective of the $f_{op}$. 
Thus, we can safely say that Signal $\textrm{I\hspace{-.1em}V}$ is for real and stems from the Al nuclei in the 20-nm layer that feel the local magnetic field generated by the Ni$_{45}$Fe$_{55}$ micro-magnet.

For further confirmation, we show that the intensity of Signal $\textrm{I\hspace{-.1em}V}$ and that of the Al(ref) are of the same order. 
Using the Al lattice constant of 0.405 nm, the number of Al nuclei in the 20-nm layer and that in the Al(ref) is obtained to be $7.2\times10^{17}$ and $2.9\times10^{19}$, respectively. 
Taking into account that the RF penetrates into the Al(ref) from both sides with the skin depth of about 600 nm, only $2\times600$ nm/10 $\mu$m of the $2.9\times10^{19}$ nuclei can be detected. 
As a result, the integrated intensity from the 20-nm layer to that in the Al(ref) should be $7.2\times10^{17}:3.5\times10^{18} =1:4.9$.

Finally, we briefly comment on the \textit{gradient} of the $B_{local}$. For this purpose, we need to take into account that the Ni$_{45}$Fe$_{55}$ was sputtered also on the "bottom" of the ditches. Our preliminary simulation showed that the line shape of Signal $\textrm{I\hspace{-.1em}V}$ can be well-reproduced when a field gradient of $\sim 0.38$ T/$\mu$m is assumed. 

In conclusion, using inductive NMR method, we have succeeded in directly observing the local magnetic field $B_{local}$ generated by the micro-magnet. The success is based on our sensitivity sufficiently high to detect nuclear-spin signals from the 20-nm layer as well as on the well-fabricated sample. From the analysis of the spectra, we found that the $B_{local}$ was $+0.17\pm0.02$ Tesla ($-0.20\pm0.01$ Tesla) when the $B_{ext}$ was applied perpendicular (parallel) to the Al layer. Our preliminary simulation showed that the spectra can be well-reproduced with a field gradient of $\sim 0.38$ T/$\mu$m, which needs to be tested in our future experiments. Since the sample we used is planar-shaped in which  a magnetic thin film is deposited, the present result gives a potential key element toward realizing high-resolution MRI. 

S. S. and Y. M. were supported by Japan Science and Technology Corporation, and also by Grant-in-Aid for Scientific Research (C) 19540362.

\end{document}